\documentclass[aps,prr,twocolumn,superscriptaddress]{revtex4-1}
\usepackage{amsmath,bbm,amssymb}
\usepackage{xcolor,graphics,graphicx}
\usepackage[export]{adjustbox}

\usepackage{subfigure}

\usepackage{times}
\usepackage{qcircuit}
\usepackage{natbib}
\usepackage{hyperref}
\usepackage{braket}
\hypersetup{
colorlinks=true,
linkcolor=blue,
citecolor=blue,
filecolor=magenta,
urlcolor=cyan
}

\usepackage[capitalize]{cleveref}

\usepackage[utf8]{inputenc}

\newcommand{\be}{\begin{equation}}
\newcommand{\ee}{\end{equation}}
\newcommand{\id}{\mathbbm{1}}

\newcommand{\tr}{\text{Tr}}
\newcommand{\bmat}{\begin{pmatrix}}
\newcommand{\emat}{\end{pmatrix}}

\begin{document}

\title{Unified collision model of coherent and measurement-based quantum feedback
}
\author{Alfred Harwood}
\affiliation{Department of Physics \& Astronomy, University College London, Gower Street, WC1E 6BT, London, United Kingdom}
\author{Matteo Brunelli}
\affiliation{Department of Physics, University of Basel, Klingelbergstrasse 82, 4056 Basel, Switzerland}
\author{Alessio Serafini}
\affiliation{Department of Physics \& Astronomy, University College London, Gower Street, WC1E 6BT, London, United Kingdom}

\begin{abstract}
We introduce a general framework, based on collision models and discrete CP-maps, to describe on an equal footing coherent and measurement-based feedback control of quantum mechanical systems. We apply our framework to prominent tasks in quantum control, ranging from cooling to Hamiltonian control. Unlike other proposed comparisons, where coherent feedback always proves superior, we find that either measurements or coherent manipulations of the controller can be advantageous depending on the task at hand.  Measurement-based feedback is typically superior in cooling, whilst coherent feedback is better at assisting quantum operations. Furthermore, we show that both coherent and measurement-based feedback loops allow one to simulate arbitrary Hamiltonian evolutions, and discuss their respective effectiveness in this regard.
\end{abstract}

\maketitle

\noindent{\em Introduction.---}The designation `coherent feedback' (CF) is used to describe a broad class of quantum control strategies where the controller is itself a quantum system which processes quantum information. It can be contrasted with `measurement-based feedback' (MF), in which the controller processes classical information resulting from measurement outcomes.  
The theory of MF goes back to the work of Belavkin, who first extended ideas from classical control theory to quantum systems  \cite{belavkin2004towards}. CF has instead its origins in the `all-optical feedback' and `feedback without measurements' \cite{wiseman1994all, wiseman1994quantum}, based on the input-output formalism \cite{gardiner1985input}. There, the output from a cavity is used to modulate the cavity's dynamics (without measurements being performed) and the process has a clear, directional `feedback loop' structure \footnote{Later, the term `coherent quantum feedback' was used to describe more general interactions between two quantum systems aimed at a certain control task \cite{lloyd2000coherent, jacobs2014coherent}. In our work, we focus on the definition involving explicit loops, as made clear in the text.}.  MF has long been one of the key techniques for achieving quantum control \cite{smith02,sayrin11,wiseman2009quantum,serafozzi12} and CF has also emerged as a powerful alternative over the past few years, with applications in quantum optics
\cite{hamerly2012advantages,iida12,shi15}, optomechanics \cite{bennett14,wang17,coptic,woolley14,li17, naumann2014steady}, nanomechanics \cite{schmid2022coherent}, NV centres \cite{hirose16} and circuit QED setups \cite{zhang12,devoret21,crowder2021quantum}.

Both quantum feedback strategies have been widely studied theoretically, especially in the context of quantum optics \cite{wiseman2009quantum, nurdin2009coherent, yamamoto2014coherent, hamerly2012advantages, wiseman1994all, wiseman1994quantum, coptic, xiang2020static, liu2016comparing, whalen2017open}. It is often claimed that CF is inherently superior to MF, and for certain definitions, regimes and problems, this is true \cite{yamamoto2014coherent, nurdin2009coherent, hamerly2012advantages, jacobs2014coherent}. 
On the other hand, other results seem to suggest a less clearcut scenario and point at the usefulness of quantum measurements for certain tasks \cite{liu2016comparing}, which can even outperform CF, e.g. for stabilising the squeezing of a cavity mode \cite{coffee}. 
In particular, the alleged superiority of CF relies on a comparison with measurement-based unconditional, averaged dynamics without considering the 
corresponding conditional dynamics, whose stochastic jumps
cannot be reproduced by unitary means (a way to express the notorious measurement problem), or miss the potential of 
non-destructive quantum measurements, upon which system and controller evolve according to the von Neumann postulate 
and then interact with each other once again afterwards.
A broader approach is then required to fully take into account the possibilities offered by quantum measurements.

To this aim, we shall present a unified framework for describing coherent and measurement-based feedback loops as cascaded collision models. Collision models (CMs) are a class of schemes for modelling open quantum dynamics where the system repeatedly interacts with an environment, which is refreshed after each interaction \cite{rau1963relaxation, bruneau2014repeated, scarani2002thermalizing,ciccarello2021quantum, campbell21}. They have found widespread application, ranging from quantum thermodynamics \cite{strasberg2017quantum, lorenzo2015heat, lorenzo2015landauer, seah2019nonequilibrium}, to quantum optics \cite{ciccarello2017collision} and non-Markovian quantum systems \cite{giovannetti2012master, rybar2012simulation, ciccarello2013collision, kretschmer2016collision}. We should mention that a discretised model for coherent feedback, encompassing an exhaustive treatment of delays, was also proposed in \cite{whalen2017open}.

In this work we treat the environment of the collision model as a `controller' which undergoes \emph{two} sequential interactions with the system before being refreshed. In between these two interactions, the controller is processed, either through measurements (in the case of MF), or coherently (in the case of CF), a key restriction being that the system cannot be manipulated directly, but only indirectly through the controller. This sequence gives the model the directional structure of a feedback loop. By assuming that the controller is always refreshed to the same state after each iteration and by keeping the system-environment interaction the same in both cases, we are ensuring that the only extra resource that MF has access to is the ability to perform measurements. We therefore argue that this framework provides a fair comparison between MF and CF.  
If the system is subject to additional noise, we assume that the controller does not have access to its source (as would be the case in quantum feed\emph{forward}---a distinction made in \cite{jacobs2014coherent}). 

We use our framework to investigate the efficacy of MF and CF for relevant examples of control tasks and uncover a highly non-trivial relationship between the two feedback strategies, depending on the task at hand. Control tasks can broadly be grouped into two classes: \emph{state control}---where the goal is to prepare the system in a specific state---and  \emph{operator control}---where the goal is to simulate the effect of a particular CP-map (often a unitary) without knowledge of the input state \cite{jurdjevic1972control, schirmer2003controllability, albertini2003notions, d2021introduction, dong2010quantum,schirmer2002quantum}. Through analytic optimisation of some of these tasks, we  show that MF is often better at achieving state stabilisation and control, especially when the system is subject to noise, whereas CF tends to outperform MF in operator control, where the preservation of quantum coherence in all bases is key. 
As examples of state control, we consider the steady-state cooling of a $d$-level quantum system (qudit) subject to noise and the stabilisation of a qubit in an excited state, counteracting decay. For operator control, we investigate the archetypal task of implementing a bit-flip on an unknown pure state qubit input. We also investigate the task of complete unitary controllability (being able to generate \emph{any} unitary evolution) in the limit of weak system-controller interaction.  \medskip

\begin{figure}[t!]
\includegraphics[width={1\columnwidth}]{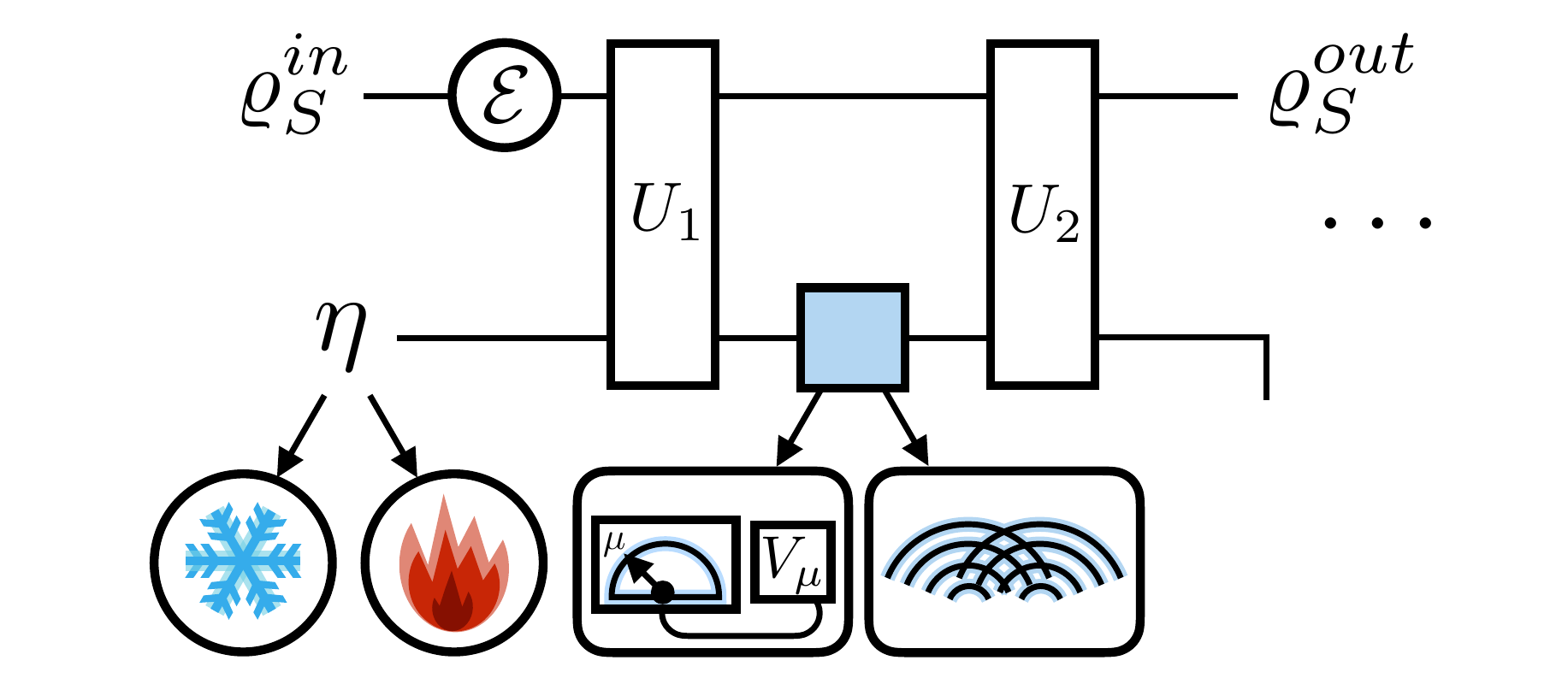}
\caption{A schematic diagram for a single ``feedback collision". A system is subject to noise channel $\mathcal{E}$ and then interacts twice with a controller. In between these collisions, either measurement-plus-unitary or just a unitary are implemented on the controller, thus realising measurement-based feedback (MF) or coherent feedback (CF), respectively. The feedback dynamics is obtained by repeated iterations of this circuit. We include both clean (``cold'') and noisy (``hot'') controllers, initialised respectively in pure or mixed states. \label{collofig1}}
\end{figure}

\noindent {\em A Collision Model of Quantum Feedback.---}We start by outlining a common framework encompassing both MF and CF, as sketched in Fig.~\ref{collofig1}. CF is described by iterated cycles each comprising the following sequence of events: i) The system is subject to noise given by a CP-map $\mathcal{E}$, representing any internal dynamics and inaccessible noise;
ii) The system and controller interact through a unitary $U_1$; iii) A unitary operation, chosen from a set $\{V_j\}$ 
(constrained, in practice, by experimental limitations), is performed on the controller. 
Here, we expand the definition of the controller to include the subsystems which interact directly with the main system, 
as well as any other additional subsystems; iv) The system and controller interact again, through another unitary $U_2$; v) The state of the controller is refreshed to its initial state $\eta$, so that it retains no memory of the previous interaction. 

MF is implemented in the same way, with step iii) being replaced with the following step: iii$^\prime$) A non-destructive measurement, described by a POVM, is performed on the controller (the available measurements being determined by experimental limitations). After the measurement, depending on the measurement result $\mu$, a unitary $V_\mu$, from the set $\{V_j\}$ is performed on the controller and available ancillas. 
%
%
%

This model allows for a fair comparison between MF and CF since the system noise, controller state and system-controller couplings are kept the same in both cases, and so is the set of unitaries that can be applied to the controller. The only difference between the two cases is the presence or absence of an in-loop measurement. Thus, comparing these two classes allows us to ask `does performing a measurement on the controller act as a resource for a particular task, or is it a hindrance?'. Indeed, in a sense that is made clear in the Supplemental Material (SM), this model allows CF to be framed as MF in the limit of infinitely weak measurements, putting both strategies on different ends of the same spectrum.

Typically, in MF schemes, the output system state will depend on the measurement record. If the measurement record is not kept, then the system will evolve deterministically to an ``unconditional'' state resulting from averaging over the measurement outcomes. If the measurement record is kept, then the ``conditional'' system state will typically evolve stochastically (although there exist cases where the subsequent operations can undo the measurements' stochasticity).  \medskip

\noindent {\em Feedback for cooling I: Noisy Controller.---}As a first case study, we apply our framework to the pivotal task of feedback cooling. We take the system and controller to have the same Hilbert space dimension $d$ and both couplings $U_1$ and $U_2$ to be partial swaps, as commonly used in CMs \cite{ciccarello2021quantum, mccloskey2014, campbell2019precursors}, i.e., we set $U_s = \sqrt{\tau} \id - i \sqrt{1-\tau} \hat{S}$, where $\hat{S}$ is the unitary swap operator and $0\le\tau\le1$ is the transmissivity; this choice of system-controller interaction will be used for the remainder of this work. We assume that the noise $\mathcal{E}$, takes the form of a depolarising channel acting on the system as 
$\mathcal{E}(\rho) = \lambda \rho + (1-\lambda) \frac{1}{d} \id $
where $\lambda$ determines the strength of the noise. In this section, we consider the case where the  controller is in the maximally mixed---or infinite temperature---state $\eta=\frac{1}{d}\id$, which we refer to as ``noisy controller''. 
The relevant figure of merit is the entropy of the steady state of the system (i.e., a state which is unchanged by the application of a single iteration of the MF or CF protocol). If the steady state exists and is unique (as will be the case in our examples), 
any input state will tend to the steady state upon repeated applications of the protocol.

For CF, the set of allowed in-loop operations on the controller is the set of single qudit unitaries. 
It is shown in the SM, through a general argument hinging on the subadditivity of entropy that, regardless of the in-loop unitary used, CF is incapable of counteracting the noise on the system and leads to a steady state which is the maximally mixed state $\rho_S = \frac{1}{d}\id$. 

For MF, 
we consider a protocol with an in-loop projective measurement in an arbitrary basis $\{\ket{j},j=0,\ldots,d-1\}$. After measurement, a unitary is applied to the controller which maps all post-measurement states to the same state, which we will label $\ket{0}$. Considering the averaged dynamics of this MF protocol, the system reaches the steady state
\be
    \rho_S = \frac1d \frac{d(1-\tau) + \tau - \lambda \tau^2}{1-\lambda \tau^2} \ket{0}\bra{0} + \sum_{j=1}^{d-1}\frac1d \frac{\tau-\lambda \tau^2}{1-\lambda \tau^2} \ket{j}\bra{j} \, ,
\ee
which is clearly lower entropy than the maximally mixed state for all values of the parameters except for the trivial case where $\tau=1$ and the system and controller do not interact. Thus, we have shown that that introducing projective measurements into the feedback loop leads to lower system entropies.  If a cold environment (in the form of a pure controller) is not available as a resource, MF outperforms all possible CF protocols. In this case measurements act as a powerful tool for preparing the controller's state, which cannot be reproduced through unitary means. 

\begin{figure}[t!]
\includegraphics[width={0.9\columnwidth}]{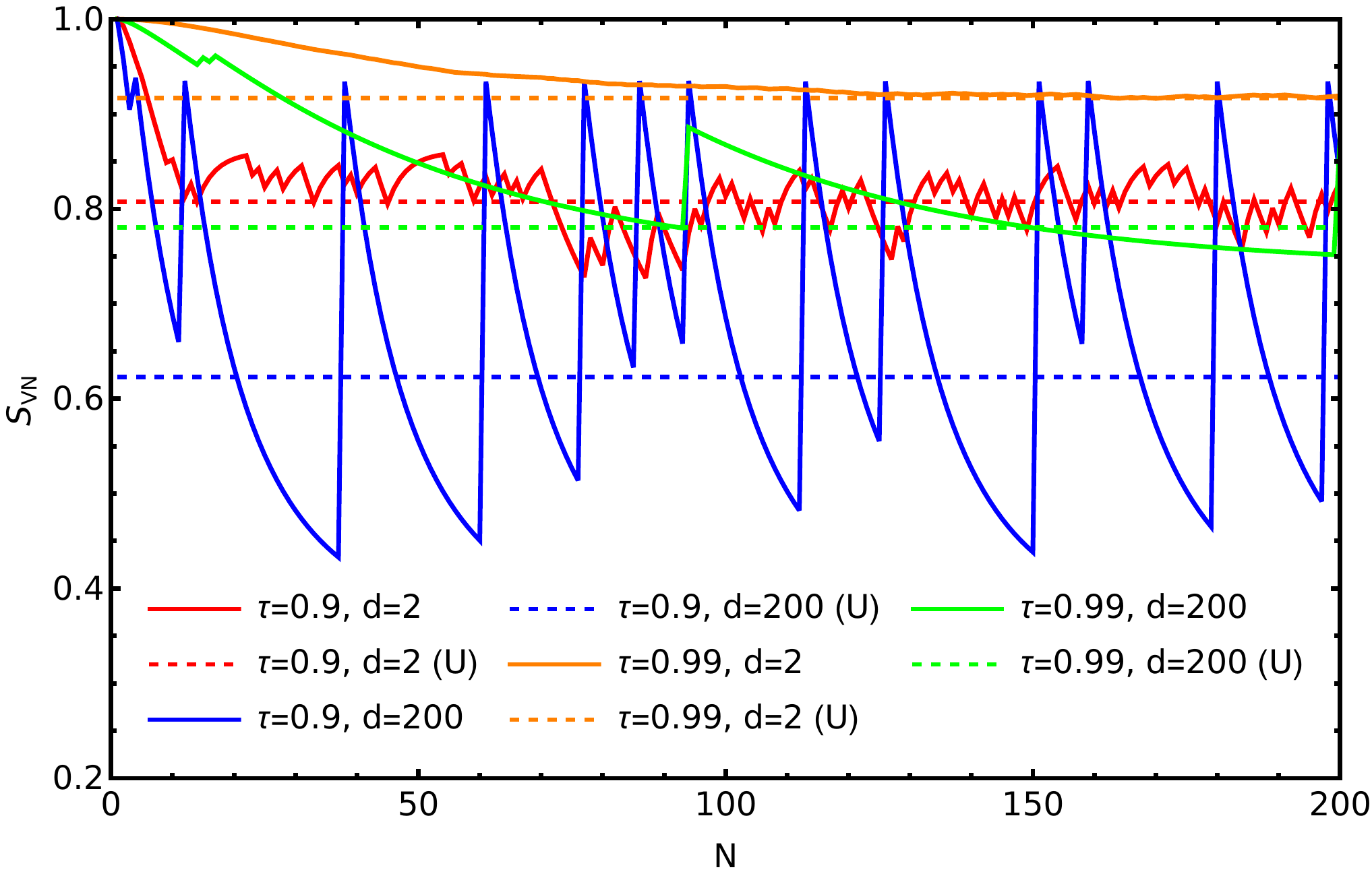}
\caption{Normalised (with base-$d$ logarithms) von Neumann entropy of conditional (continuous lines) and steady-state unconditional (horizontal dotted lines) measurement-based feedback cooling for different values of $\tau$, $\lambda$ and $d$, for a maximally mixed initial state. The $x$-axis reports the number of iterations. \label{Stofig1}}
\end{figure}

We now consider the conditional dynamics of this MF protocol. In general, the stochasticity of measurement outcomes, unless averaged upon, does not yield a steady state and instead results in oscillating entropies as shown in Fig.~\ref{Stofig1}. However, in the case of an ideal full swap ($\tau=0$) the post-measurement unitary can completely undo the stochasticity of the measurement, leading to the  steady state $\ket{0}$, i.e., perfect cooling.
Examining Fig.~\ref{Stofig1} one can see that the asymptotic cooling performance, in terms of steady state entropy, decreases with $\tau$ (as lower $\tau$ imply the possibility of injecting 
low entropy states from the controller) and, as one should expect, decreases 
with the noise strength $\lambda$. One may also see that the entropy `jumps' become larger for higher system dimensions.
Besides, this study also allows us to estimate the cooling rate of the model, i.e., the typical number of iterations needed to approach 
the unconditional entropy, as $1/(1-\tau)$ (in that the interaction with the controller dominates the cooling process).

By applying a quantum entropy power inequality \cite{Audenaert16,Carlen16}, one may show that, for an input diagonal in the measurement basis, 
the entropy of the conditional output state is minimised when, after measuring the controller, all measurement outcomes are mapped to the system's dominant eigenvector (corresponding to the system's largest eigenvalue). Furthermore, when the input is diagonal in the measurement basis, the output is diagonal in the same basis, with the same dominant eigenvector. Thus, for repeated application, starting with maximally mixed input state or any input state with dominant eigenvector $\ket{0}$, our MF protocol is optimal when considering conditional dynamics.  \medskip


\noindent {\em Feedback for cooling II: Clean Controller.---}While MF outperforms CF when cooling with a maximally mixed controller, this is not always the case when the controller is in a pure state. 
Here we consider the controller to be in the pure state $\eta=\ket{0}\bra{0}$, which we refer to as a ``clean controller''. 
In the case of CF, no in-loop unitary is needed, because the system is swapped with the pure controller, thereby getting cooled.
In the SM, we prove analytically that this CF loop is optimal for qubits assuming $\ket{0}$ is an eigenvector of the steady state. 
Here, we present the linear entropies for qubit systems in view of their more compact expressions (more general expressions for arbitrary dimension can be found in the SM). 
The steady-state linear entropy for the MF protocol is $ S_{MF} = \frac{1}{2}- \frac{\left(\tau^2-1\right)^2}{2 \left(\tau^2 \lambda-1\right)^2} $, while for CF protocol we have $S_{CF}=\frac{1}{2}-\frac{8 \tau^2 \left(\tau-1\right)^2}{\left(\left(1-2\tau\right)^2 \lambda-1\right)^2}$.
Comparing the two expressions, we find that $S_{MF}<S_{CF}$ for $\tau<1/3$ and $S_{MF}>S_{CF}$ for $\tau>1/3$. In other words, there are two regimes: one where MF outperforms CF and one where CF outperforms MF. 
Note that when $\tau=1/2$, $S_{CF}=0$ and CF is able to stabilise a pure state. On the other hand, no MF protocol with projective measurements can stabilise a pure state for $\tau=1/2$ (see SM). Conversely, for $\tau=0$, MF is always capable of stabilising a pure state using the protocol described earlier, 
whilst CF cannot stabilise any state other than the maximally mixed state, as the action of a unitary alone cannot undo the effect of the depolarising map. A comparison of MF and CF for cooling at intermediate controller noises (where the controller is neither pure nor maximally mixed) can be found in the SM.  \medskip


\noindent {\em Protecting an excited state from decaying.---}After a task based on entropy, we now consider one focused on the system's energy, namely protecting an excited state from environment-induced decay. For simplicity, we consider a qubit initially prepared in the excited state $\ket{1}$ (with $\ket{0}$ being the ground state) subject to decay, modelled by $\mathcal{E}$ with Kraus operators $E_0= \sqrt{\gamma}\ket{0}\bra{1}, \; E_1 =\sqrt{1-\gamma}\ket{1}\bra{1} + \ket{0}\bra{0}$. A feedback loop, either CF or MF, is employed to counter the effect of decay. We take the controller to be noisy. Our figure of merit is then the steady-state occupation of the $\ket{1}$ state.
For CF, we restrict the in-loop unitaries to rotations $U=\cos \chi\id+i\sin\chi \sigma_y$.
and assume that the controller is initialised in the maximally mixed state. Investigating numerically, we find that the optimal CF protocol depends on the partial swap strength; for $\tau>\frac{1}{2}$ $(\tau<\frac{1}{2})$, the optimal protocol involves setting $\chi=0$ $(\chi=\frac{\pi}{2})$. 

For $\tau>\frac{1}{2}$, increasing the noise strength decreases the steady-state occupation of the excited state, as expected. However, for strong system-controller interactions $\tau<\frac{1}{2}$, the optimal performance (when $\chi=\frac{\pi}{2}$) is actually \emph{improved} by stronger damping. This is because the action of the in-loop $\frac{\pi}{2}$ rotation is more effective at populating the $\ket{1}$ state when more of the state is initially prepared in the $\ket{0}$ state. In this sense, CF allows for the purifying effect of the amplitude damping map to be harnessed for the purpose of increasing the excited state population.
When $\chi=0$ and the in-loop unitary is the identity, the steady-state occupation of the excited state is
\be \label{AD:X0}
    \rho_{11}^{\chi=0} = \frac{2 \tau \left(1-\tau\right)}{4 \left(\tau-1\right) (\gamma -1) \tau+\gamma }, 
\ee
which is a decreasing function of $\gamma$ for $\tau>\frac{1}{2}$. Conversely, when $\chi=\frac{\pi}{2}$, the steady-state occupation is
\be \label{AD:Xpi2}
    \rho_{11}^{\chi=\frac{\pi}{2}} = \frac{1-\tau}{2 (\gamma -1) \tau-\gamma +2} \, ,
\ee
which is an increasing function of $\gamma$ for $\tau<\frac{1}{2}$. 

\begin{figure}
    \centering
    \includegraphics[width={0.9\columnwidth}]{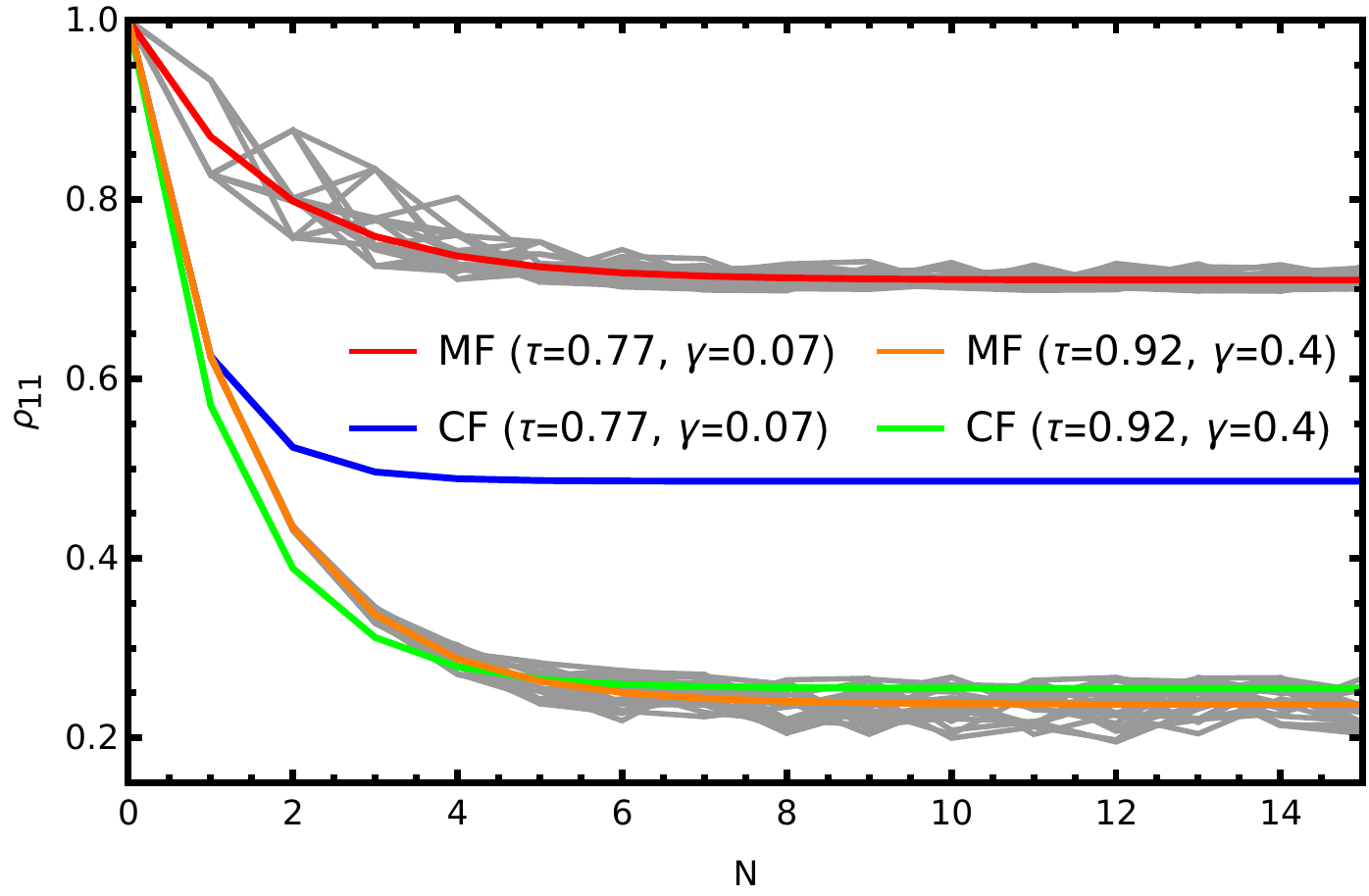}
    \caption{Occupation of the excited state against number of iterations of MF or CF. The MF trajectories shown in red and green are the unconditional trajectories. For each setup fifty conditional, filtered trajectories are shown in light grey.}
    \label{fig:r00vsN}
\end{figure}

We compare this to an intuitive MF protocol which measures the controller in the $\{\ket{1}, \ket{0}\}$ basis, does nothing if the result is $\ket{1}$, and applies an in-loop rotation with $\chi=\frac{\pi}{2}$ if the result is $\ket{0}$. This protocol results in an excited steady-state occupation 
\be \label{AD:mf}
    \rho_{11}^{MF}=
    \frac{2-\tau^2-\tau}{2 (\gamma -1) \tau^2+2} \, .
\ee
This expression is greater than equation (\ref{AD:Xpi2}) for all values of $\tau$ and $\gamma$, so that MF outperforms CF in the regime of strong system-controller coupling $\tau<\frac{1}{2}$. However, for  $\tau>\frac{1}{2}$, equation (\ref{AD:mf}) is slightly lower than (\ref{AD:X0}). In particular, (\ref{AD:X0}) is greater than (\ref{AD:mf}) when $\tau>\frac{1}{2} \frac{-7 \gamma +\sqrt{\gamma  (17 \gamma -24)+16}+4}{2-2 \gamma }$.
In this regime, CF will outperform MF, though numerical investigations suggest that the advantage is small. Conversely, in the regime where MF outperforms CF, the advantage tends to be larger. Figure (\ref{fig:r00vsN}) shows $\rho_{11}$ the occupation of the excited state, against the number of iterations of the feedback loop. Both MF and CF protocols described above are presented. The unconditional (averaged) MF trajectory is shown, as well as fifty conditional trajectories for each setup. \medskip

\noindent{\em Operator control: Performing a bit flip operation.}---As a first paradigmatic task in operator control, we shall consider the implementation of a bit flip operation on a qubit system ($d=2$); we stress that the bit flip must be induced on the system by only acting on the controller. We allow in-loop operations to be any single qubit unitary, we assume that the system is not subject to any noise (i.e., the map $\mathcal{E}$ in Fig.~\ref{collofig1} is the identity) and that the controller is initialised to a generic mixed state $\eta=\eta_{0}\ket{0}\bra{0}+(1-\eta_0)\ket{1}\bra{1}$. The system input state is of the general form $\ket{\psi} = \cos{\frac{\chi}{2}}\ket{0} + e^{i \phi}\sin{\frac{\chi}{2}}\ket{1}$ and the figure of merit will be the Haar-averaged fidelity of the output state to the desired state, i.e., $A=(4\pi)^{-1}\int \mathrm{d}\phi \mathrm{d}\chi \sin\chi \bra{\psi_X}\rho_S^{out}\ket{\psi_X}$,
where $\ket{\psi_X} = \sigma_x \ket{\psi}$. Crucially,  the feedback protocol must be implemented without knowledge of the exact input state (i.e., without knowledge of $\chi$ and $\phi$). 

First, we consider an intuitive CF protocol, where the in-loop operation performed is a $\sigma_x$ unitary. When the state $\ket{\psi}$ is put through this protocol, 
we obtain the average fidelity for coherent feedback (see SM)
\be
    A_{CF} = 1-\frac{2}{3}\tau\, .
\ee
We will now show that no MF protocol involving projective measurements can outperform this CF protocol. Our general MF protocol will be as follows. After interacting through the partial swap, the controller will be measured in the $\{\ket{0}, \ket{1}\}$ basis (notice that all choices of basis are equivalent due to the unitary covariance of bit-flips and of the Haar measure). If the measurement result is $\ket{0}$, a general unitary $U$ will be applied to the controller and if the result is $\ket{1}$, a different unitary denoted by $V$ will be applied instead. After the system and controller interact again through the partial swap, the Haar measure averaged fidelity to the desired output state for this general MF protocol is a function of $U$ and $V$. It is shown in the SM that this fidelity is maximised when $U=V=\sigma_x$, which yields the following output fidelity for the optimal MF protocol
\be
    A_{MF} = \frac{2}{3} - \frac{1}{3}\tau.
\ee
It is simple to see that this fidelity is always smaller than that achieved through CF (except for the trivial case $\tau=1$). Furthermore, we find that the performance of MF improves with (non-projective) weaker measurements, but cannot outperform CF for \emph{any} in-loop POVM (see SM). Therefore, CF is  superior to MF in assisting bit-flipping. \medskip

\noindent{\em Operator control in the limit of weak interactions.---}Having considered the task of implementing a specific unitary, we now investigate the general task of implementing an arbitrary unitary evolution on the system. To this aim, we consider the setup discussed above in the limit of infinitesimally weak system-controller interactions, i.e.,  
we let $\theta\xrightarrow[]{}\text{d}\theta$ in the transmissivity of the partial swap $\tau=\cos^2 \theta $. Expanding the interaction and keeping only first-order terms, we find that  one iteration of MF or CF results in the following transformation on the system density operator:
\be
    \rho_S \xrightarrow[]{} \rho_S -i[(\Lambda_j(\eta)+\eta), \rho_S] \text{d}\theta ,
\ee
where $\Lambda_j$ represents the particular CP-map applied in-loop (unitary for CF and measurement followed by unitary feedback for MF) from the set of possible in-loop CP-maps $\{\Lambda_j\}$. 
Therefore, in this weak limit, the system evolves unitarily for both MF and CF, under a unitary given by $V_j = e^{-i H_j \text{d}t}$ where $H_j =(\Lambda_j(\eta)+\eta)$. By iteratively applying this process, both MF and CF can simulate any Hamiltonian on the system, provided that it falls within the algebra generated by commutation of elements of the set $\{\Lambda_j(\eta)+\eta\}$ (assuming that the $\Lambda_j$ can be changed after each iteration as necessary). 
If the set of in-loop CP-maps generated by MF or CF is non-trivial, this algebra is typically the entire space of Hermitian matrices, since all pairs of Hermitian matrices, except a set of measure zero can generate the entire space of Hermitian matrices by commutation \cite{lloyd1995almost}. Thus, in the limit of weak interactions, as long as the set $\{\Lambda_j(\eta)+\eta\}$ is non-trivial, both MF and CF can be used to simulate the effect of any Hamiltonian on the system and for this purpose are equivalent. 
However, the dramatic ability of measurements of changing the controller's spectrum can be leveraged here to achieve faster evolutions 
to target states than in CF schemes, as $\Lambda_j(\eta)$ can in principle be any quantum state in MF. 
In this regard, let us also notice that in the limiting case where $\eta$ is maximally mixed, the set $\{\Lambda_j(\eta)+\eta\}$ for CF will contain only one element which will be proportional to the identity (since the action of any unitary leaves the identity unchanged), and 
so CF will not be able to exert any Hamiltonian control.\medskip

\noindent{\em Conclusion.---}We have introduced a common framework for coherent and measurement-based quantum feedback based on a collision model, featuring repeated interactions with an environment which acts as a controller. 
This allows MF and CF to be compared on equal footing, under the assumption of non-destructive measurements.
Our framework is broadly applicable to practical cases, ranging from quantum optics and optomechanics in the continuous variable regime to linear optics \cite{santos15,marletto20}, extending to all conceptual situations where CMs find application \cite{campbell21}. We note that, since the input-output formalism can be framed as a CM \cite{ciccarello2017collision}, if the continuous-time limit of this model is taken it fully captures the quantum optical notion of coherent feedback as described in, e.g.~\cite{coffee, coptic} or in the equivalent SLH formulation \cite{nurdin2009coherent,nurdingough}.


The case studies addressed  provide one with strong indications of the general advantages and disadvantages of each type of feedback.


\noindent{\em Acknowledgements.---}
M.B. acknowledges funding from the Swiss National Science Foundation (PCEFP2$\_$194268).

\bibliography{ColloBib.bib}

\widetext

\begin{center}{\bf Supplemental Material}\end{center}
\section{Coherent Feedback Cooling with noisy controller}
We prove here that for `noisy' controllers initialised in the maximally mixed state, coherent feedback cannot provide any advantage.
All of our examples use the model introduced in the paper. In this model, the system and controller are both $d$-dimensional qudits which interact twice through a partial swap unitary. Inside the coherent feedback loop, we have no ancillary systems, and the only allowed in-loop operations are single qudit unitaries. Throughout these notes, the subscript $T$ will indicate the total two-qudit, system-controller joint state and the subscript $S$ will refer to the system alone. Our setup for cooling with coherent feedback at high temperature unfolds as follows:
\begin{enumerate}
    \item The controller is initialised to the maximally mixed state $\eta^{in} = \frac{1}{d}\id$. The system input $\rho_S^{in}$ is the result of the previous iteration of the process.
    \item The system is subject to noise in the form of a depolarising channel.
    \item The system and controller interact through a partial swap unitary.
    \item A unitary $U$ is applied to the controller.
    \item The system and controller interact again through a partial swap unitary.
\end{enumerate}
The von Neumann entropy of the total input state $\rho_{S}^{in} \otimes \eta$ is given by:
\be
    S_T^{in} = S(\rho_{S}^{in} \otimes \eta) = S(\rho_{S}^{in}) + S(\eta^{in}) \, .
\ee
After the application of the noise, the system is still in a separable state and the total entropy is:
\be
    S_T = S(\mathcal{E}(\rho_S^{in})) + S(\eta^{in}) \, .
\ee
Since the rest of the protocol can be described by unitary operations, the global entropy is left unchanged so
\be
    S_T^{out} = S_T = S(\mathcal{E}(\rho_S^{in})) + S(\eta^{in}) \, .
\ee
Using the subadditivy of entropy, we can obtain the following bound on the entropy of the system output state:
\be
    S(\rho_S^{out}) \geq S_T^{out} - S(\eta^{out}) = S(\mathcal{E}(\rho_S^{in})) - \Delta S(\eta) \, .
\ee
where $\Delta S(\eta) = S(\eta^{out}) - S(\eta^{in})$ is the change in entropy of the controller. Since $\eta^{in}$ is maximally mixed, the maximum value of $\Delta S(\eta^{in})$ is $0$. This gives us the bound on the entropy of the output:
\be
    S(\rho_S^{out})\geq S(\mathcal{E}(\rho_S^{in}))\geq S(\rho_S^{in}) \, .
\ee
The entropy of the system output state can never be lower than the entropy of the input. Furthermore, the input and output entropies are only equal when $S(\mathcal{E}(\rho_S^{in}))= S(\rho_S^{in})$ which is only true when $\rho_S^{in}$ is maximally mixed, meaning that the only steady state is the maximally mixed state.

\section{ Measurement-Based Feedback for Cooling with a noisy controller}
In this section we consider measurement-based feedback cooling for a two-qudit setup with an initially maximally mixed environment/controller $\eta =\frac{1}{d}\id$. The system undergoes a depolarising map, before system and controller interact through a first partial swap. The controller is then measured in the basis $\{\ket{j}\}$ and a unitary $V_j$ is applied to the controller, depending on the measurement outcome. The system and controller then interact again through a partial swap, and the controller is traced over. Throughout the supplemental material, we use the convention that the partial swap is written $U_s = c \id -i s \hat{S}$ where $c=\cos \theta$ and $s=\sin \theta$, in contrast to the main paper, where it is written in terms of the transmissivity $\tau = c^2$.
\subsection{Unconditional MF}
We will now derive the steady state for the case of measurement-based feedback averaged over all trajectories.  This means that the in-loop measurement and feedback can be represented as a CP-map with elements $\{V_j\ket{j}\bra{j}\}$. In particular, we will find the steady state for the case where, through the action of $V_j$ , all measurement outcomes are mapped to the same state, which we will label $\ket{0}$. This corresponds to a CP-map with elements $\{\ket{0}\bra{j}\}$. In the next section we will show that this is optimal when one filters and conditions the state on the measurement outcome. We will first derive expressions for the most general projective MF protocol, and then restrict to the optimal case. Throughout, we will denote the action of the initial depolarising map as $\rho_N=\mathcal{E}(\rho_{in})=\lambda \rho_{in} + \frac{1-\lambda}{d}\id$. Notice that $\rho_N$ has the same eigenvectors as $\rho_{in}$ as well as the same ordering (ie. the eigenvector corresponding to the largest eigenvalue of $\rho_{in}$ also corresponds to the largest eigenvalue of $\rho_N$).

After the action of the depolarising map and the first partial swap, the system and controller are correlated in the following state:
\be
    \rho_T = c^2 \rho_N\otimes \eta + s^2 \eta \otimes \rho_N - i s c [S, \rho_N\otimes \eta] \, ,
\ee
where $S$ is the swap unitary. After the measurement, and application of the unitaries $\{V_j\}$, the global state is
\be \label{mixture}
    \rho_T=\sum_j p_j \rho_j \otimes \ket{\psi_j}\bra{\psi_j} \, ,
\ee
where $\ket{\psi_j} = V_j \ket{j}$ and $p_j$ are the probabilities of each outcome. The (un-normalised) system state for each measurement outcome is given by 
\be \label{rhoj}
    p_j \rho_j = \frac{1}{d}c^2 \rho_N + s^2 \eta (\rho_N)_{jj} -isc [\ket{j}\bra{j}, \rho_N] \,  ,
\ee
where $(\rho_N)_{jj} = \bra{j}\rho_N \ket{j}$. We have also used the fact that $\tr_C [S, A\otimes B] = [B, A]$.
After a final interaction through a partial swap, the controller is traced out, leaving the output state:
\be \label{out}
    \rho_{out} = \sum_i p_i \bigg(c^2 \rho_i + s^2 \ket{\psi_i}\bra{\psi_i} -i s c [\ket{\psi_i}\bra{\psi_i}, \rho_i]\bigg) \, .
\ee
The optimal protocol, in which all measurement results are mapped to the $\ket{0}$ state through the unitaries $V_j$, amounts to setting $\ket{\psi_j}=V_j\ket{j}=\ket{0}$ for all $j$. Using $\sum_j p_j \rho_j =c^2 \rho_N + s^2 \eta$, we find the output state:
\be 
    \rho_{out} = c^4 \rho_N + s^2 c^2 \eta + s^2 \ket{ 0}\bra{ 0} - i s c^3 [\ket{ 0}\bra{ 0}, \rho_N] \, .
\ee

The fact that $\eta=\frac{1}{d}\id$ allows us to write $\rho_N = \lambda \rho_{in} + (1-\lambda) \eta$. We now assume a steady state, which involves setting $\rho_{in}=\rho_{out}= \rho_{ss}$. Re-arranging for $\rho_{ss}$ gives:
\be
    \rho_{ss} = \frac{1}{1-c^4 \lambda} \bigg((c^4(1-\lambda) + s^2 c^2) \eta +s^2 \ket{ 0}\bra{ 0} -\lambda i s c^3 [\ket{ 0}\bra{0}, \rho_{ss}] \bigg) \, .
\ee
The only solution to this equation is one where $\rho_{ss}$ is diagonal in a basis $\{\ket{j}\}$ which contains $\ket{0}$. To see this, we can act on the above equation with $\bra{j}$ and $\ket{k}$ from the left and right (for $j\neq k$) and obtain:
\be
    \bra{j}\rho_{ss}\ket{k} = \frac{1}{1-c^4 \lambda}(- \lambda i s c ^3 (\delta_{j0}\bra{0}\rho_{ss}\ket{k} - \bra{j}\rho_{ss}\ket{0}\delta_{0k}) \, .
\ee
Note that the right hand side of this equation is only nonzero when either $j$ or $k$ is  equal to zero. For $j=0$, $k\neq0$, we obtain:
\be
    \bra{0}\rho_{ss}\ket{k} =\frac{1}{1-c^4 \lambda}(- \lambda i s c ^3 (\bra{0}\rho_{ss}\ket{k} )\, .
\ee
The only solution to this equation is when $\bra{0}\rho_{ss}\ket{k} =0$. Therefore $\rho_{ss}$ is diagonal in the basis $\{\ket{j}\}$. This means that the eigevalues of $\rho_{ss}$ are given by $\bra{j} \rho_{ss}\ket{j} $. The eigenvalue associated with $\ket{0}$ is:
\be
    \alpha_0 = \frac{1}{1-c^4 \lambda}\bigg(\frac{c^4(1-\lambda) + s^2 c^2}{d} + s^2\bigg) \, . \label{alfunc}
\ee
The remaining $d-1$ eigenvalues are degenerate, each with value:
\be
    \alpha_j  = \frac{1}{1-c^4 \lambda}\bigg(\frac{c^4(1-\lambda) + s^2 c^2}{d}\bigg) \, .
\ee
Thus, the steady state under this protocol takes the form:
\be
    \rho_{S} = \frac1d \frac{d(1-c^2) + c^2 - \lambda c^4}{1-\lambda c^4} \ket{0}\bra{0} + \sum_{j=1}^{d-1}\frac1d \frac{c^2-\lambda c^4}{1-\lambda c^4} \ket{j}\bra{j} \, .
\ee
The linear entropy for this steady state is:
\be
    S_L = 1- \sum_j \alpha_j^2 = (1-\frac{1}{d}
    ) - \frac{\left(c^2-1\right)^2 (d-1)}{d \left(c^4 \lambda -1\right)^2}.
\ee
We note that, for states like these with one large eigenvalues and $d-1$ degenerate eigenvalues, both Von Neumann and linear entropies are solely functions of the largest eigenvalue and are thus equivalent for the purposes of comparison. This is always the case for qubits, where the linear entropy and Von Neumann entropy are equivalent. Because of this, and in view of their more compact expressions, we will use the linear entropy for the remainder of this investigation (though we will give expressions for the eigenvalues of each state, from which the Von Neumann entropy can easily be calculated).
\subsection{Conditional MF}
In this section, we will consider filtered measurement feedback, meaning that the measurement result is recorded and the system's conditional state evolves stochastically. After the measurement is performed along with the in-loop unitary operation, the system and controller are in the joint state $\rho_j \otimes \ket{\psi_j}\bra{\psi_j}$ for a measurement result labelled by $j$ (where $\rho_j$ and $\ket{\psi_j}$ are defined as in the previous section). This is as opposed to averaged MF, where the system and controller are in the mixture of states given by (\ref{mixture}). After the second partial swap and tracing out of the controller, the system output will be:
\be
    \rho_{out,j} = \tr_C[U_s \rho_j \otimes \ket{\psi_j}\bra{\psi_j} U_s^\dag] \, .
\ee
We will now show that,
for a $\rho_{in}$ which is initially diagonal in the measurement basis (as would be the case for the maximally mixed state, which represents the uncontrolled steady state of the system, as well as for the unconditional steady states determined in the previous section), the entropy of this output is minimised when $\ket{\psi_j}$ is set to be the dominant eigenvector of $\rho_{in}$ (i.e., the eigenvector associated to its largest eigenvalue), regardless of the measurement result. 
Notice also that the choice of basis for the `atomic' projective measurement (i.e., a measurement that resolves each individual basis state) is irrelevant, given the 
unitary invariance of the problem (here, we are disregarding more general POVMs or adaptive measurements, 
which may become the subject of further inquiry in the future).
If $\rho_{in}$ is diagonal in the measurement basis, the commutator in equation (\ref{rhoj}) is equal to zero and we can write
\be
    \rho_j = \frac{1}{p_j}\big(\frac{1}{d}c^2\rho_N + s^2\eta(\rho_N)_{jj}\big) \, .
\ee
Note that each $\rho_j$ has the same set of eigenvectors in the same ordering as $\rho_{in}$, so the eigenvector corresponding to the largest eigenvalue is the same for both $\rho_{in}$ and $\rho_j$.

We now make use of the entropy power inequality for the output of a partial swap gate derived in \cite{Audenaert16}, in the form of the following majorisation relation:
\be
    \lambda( \rho_{out, j}) \prec c^2 \lambda(\rho_j) + (1-c^2)\lambda(\ket{\psi_j}\bra{\psi_j}) \, ,
\ee
where we have used $\lambda(\rho)$ to indicate the spectrum of $\rho$, ordered from the largest eigenvalue to the smallest. The right and left hand sides of this relation are equal when $\ket{\psi_j}$ is pointing along the direction of the dominant eigenvector of $\rho_j$, which in turn is the dominant eigenvector of $\rho_{in}$. Thus, to minimise the conditional output entropy for an input diagonal in the measurement basis, we must set all $\ket{\psi_j}$ equal to the eigenvector corresponding to the largest eigenvalue of $\rho_{in}$. 

\begin{figure}[t!]
\includegraphics[width=0.45\textwidth]{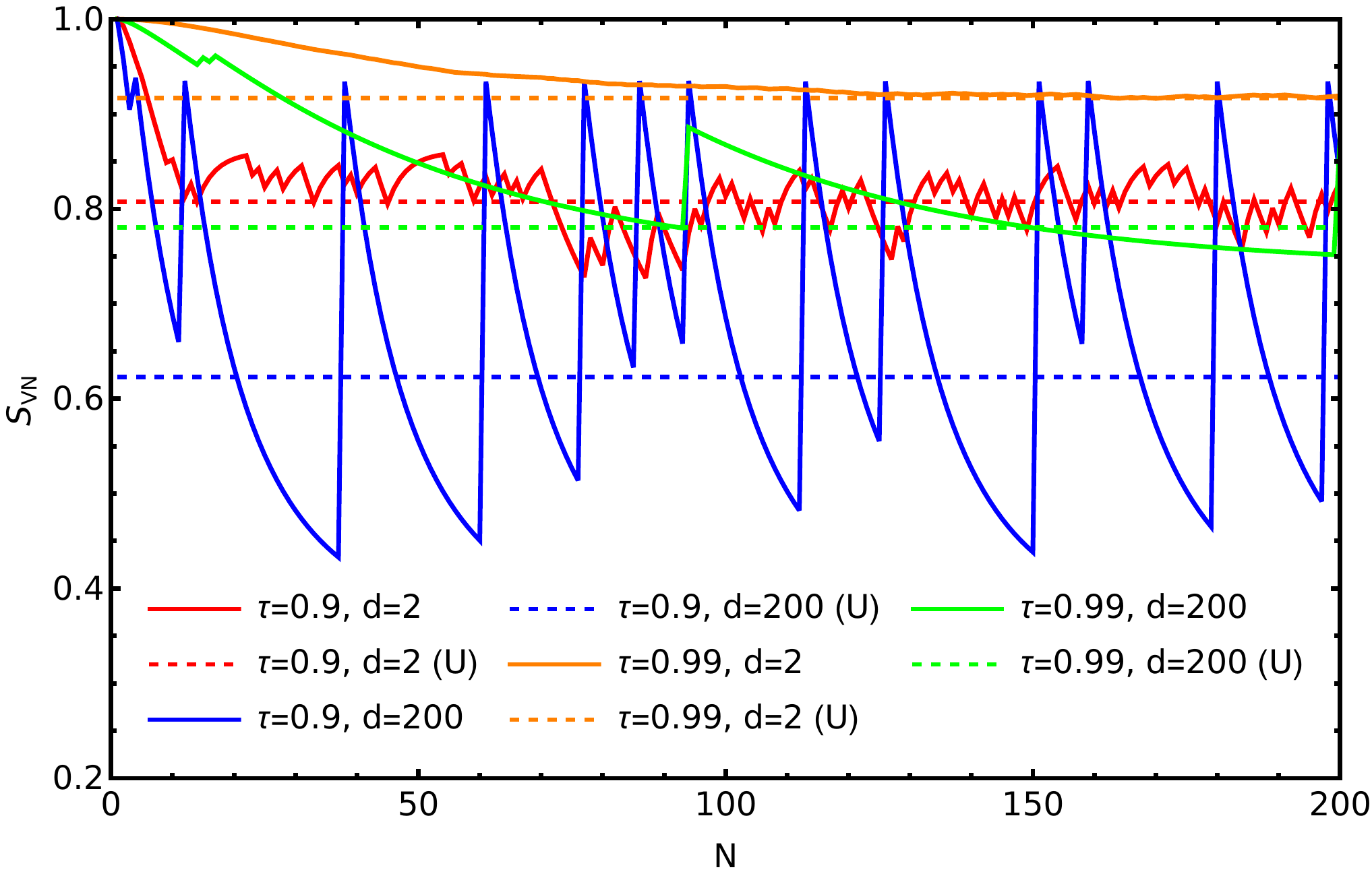}
\includegraphics[width=0.45\textwidth]{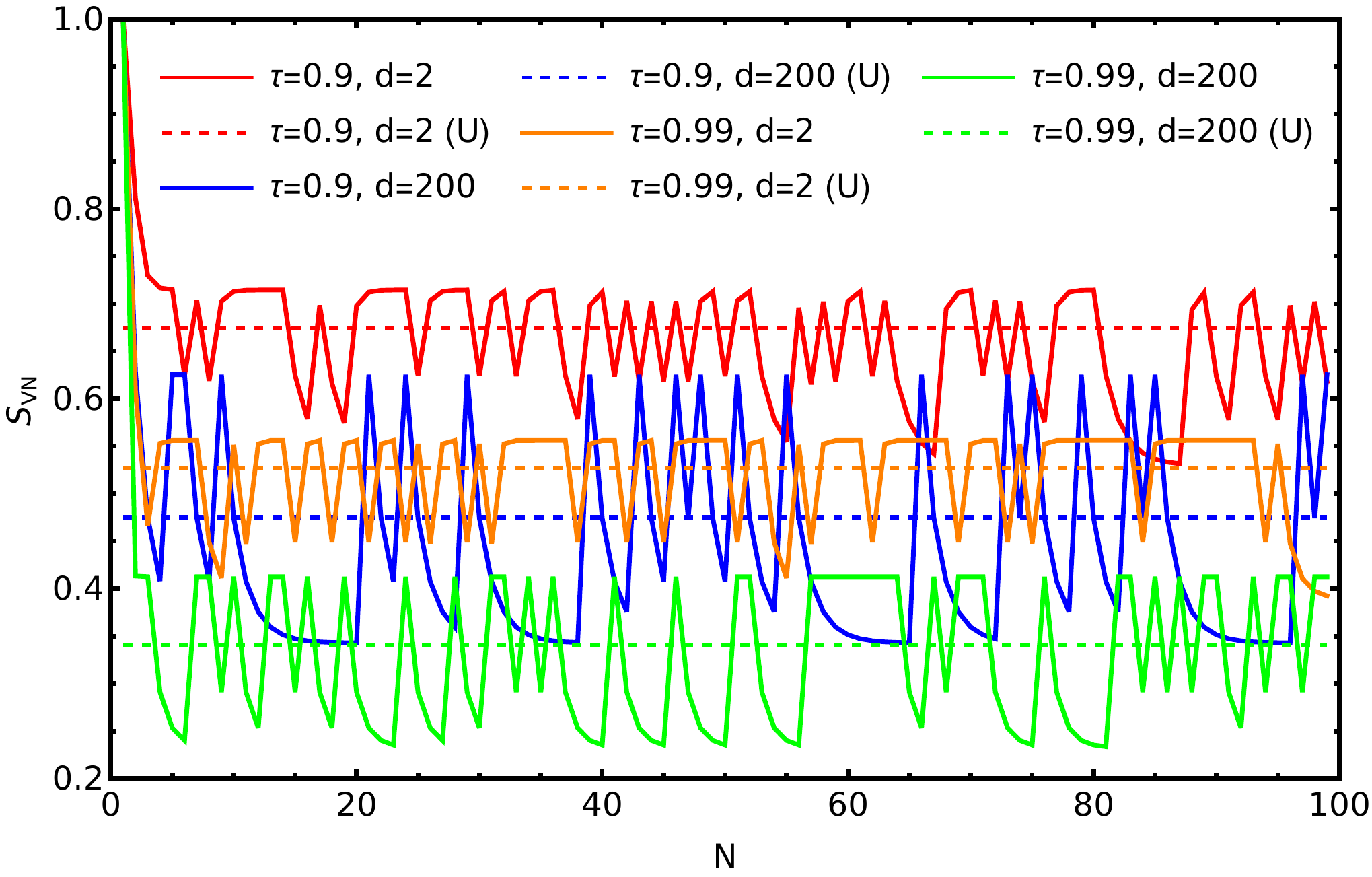}
\caption{Normalised (with base-$d$ logarithms) von Neumann entropy of conditional (continuous lines) and steady-state unconditional (horizontal dotted lines) measurement-based feedback cooling for different values of $c$, $\lambda$ and $d$, for a maximally mixed initial state. The $x$-axis reports the number of loop iterations performed. \label{stofig}}
\end{figure}

Let us now also briefly consider some quantitative examples of the conditional stochastic evolution occurring under this filtered, measurement-based feedback.
This provides insight into the dependence of the optimal cooling process on the various system parameters. 
Let us remind that the latter are the depolarising strength $\lambda$ (from complete depolarisation for $\lambda=0$ to the identity channel for $\lambda=1$), 
the partial swap angles parametrised by the `connectivity' $c=\cos\theta$ and the system dimension $d$. 
Fig.~\ref{stofig} shows the evolution of the von Neumann entropy (with base $d$ logs, such that $1$ is always maximum entropy for all dimensions) for various choices of parameters, always starting from the maximally mixed state (which corresponds to the uncontrolled steady state). 
The {\em unconditional} (`unfiltered') steady state entropies are also reported for comparison and, as may be seen, always fall within the range 
of two possible, typical unconditional values, deviating relatively little from them, 
such that the stabilised, unfiltered strategy proves to be very effective in these instances. 
Indeed, in all of these case studies, regardless of the Hilbert space dimension, only one of the system eigenvalues is different from the other, accruing probability at the expense of the other eigenvalues, which remain equal (therefore, any POVM capable of resolving the dominant eigenvector would also be optimal).
When the dominant eigenvector is detected on the controller branch of the feedback loop, its proportion in the system branch decreases, and therefore the measurement is disadvantageous in terms of cooling the system. At each run of the feedback loop, this occurs with probability
\be
p_0 = \frac{c^2}{d}+s^2 \alpha_\lambda \; ,
\ee
where $s=\sin\theta$ $\alpha_{\lambda}= \lambda\, \alpha_{in} + \frac{(1-\lambda)}{d}$ and $\alpha_{in}$ is the dominant eigenvalue of the input state.
For completeness, let us also report the values of the largest output eigenvector upon measurement of the dominant eigenvector in the feedback loop ($\alpha_{0,0}$) 
upon measurement of another outcome ($\alpha_{0,1}$), to be contrasted with the average, unconditional $\alpha_0$ of Eq.~(\ref{alfunc})
\begin{align}
\alpha_{0,0} & = \frac{c^2 \alpha_{\lambda}}{p_0 d} + s^2 \; , \\
\alpha_{0,1} & = \frac{c^2}{(1-p_0)d} \left( c^2 d \alpha_{\lambda}-\alpha_\lambda+s^2\ \right) + s^2 \; .
\end{align}
All of these three functions are monotonically increasing in $\lambda$ and decreasing in $c^2$. Therefore, as one should expect, the asymtpotic cooling performance 
will be more effective for higher $\lambda$, corresponding to less noise, and lower $c^2$, corresponding to a larger connectivity between system and controller, which 
allows one to swap a substantial part of the final state with a low entropy one. 
Hence, as shown in Fig.~\ref{stofig}, higher connectivities can offset larger noise parameters. 
Observe also that higher dimensions, typically, make for larger spread in the normalised entropy around the unconditional value.
Further, it is worthwhile noticing that the model also allows one to observe that cooling will be achieved in a number of steps of the order of $1/s^2$: the transmittivity 
between controller and system is the parameter that determines the model's cooling rate. 

\section{Coherent Feedback for Cooling with a clean controller}
We will consider a setup identical to the previous section, except that the environment is initialised to a pure state $\ket{0}\bra{0}$. We will restrict to qubits and find the lowest entropy steady state achievable through CF which is diagonal in the $\{\ket{0}, \ket{1}\}$ basis.

We express the in-loop unitary as using the general decomposition:
\be
    U =
    \bmat
        e^{i \varphi_1}\cos \chi & e^{i \varphi_2}\sin \chi \\
        -e^{-i \varphi_2}\sin \chi & e^{-i \varphi_1}\cos \chi
    \emat \, .
\ee
The steady state which satisfies the condition that it is diagonal in the basis containing $\ket{0}$ takes the form:
\be
    \rho_{ss} = \text{diag}(e_1, 1-e_1)
\ee
with
\be
   e_1= \frac{-2 c^4 (\lambda +1) p^2 (q+1)+2 c^2 \left(p^2 (\lambda 
   (q+2)+q+1)-\lambda \right)+\lambda -2 \lambda  p^2+1}{\lambda  \left(4
   c^2 \left(p^2 \left(\left(c^2-1\right) q+c^2-2\right)+1\right)+4
   p^2-2\right)-2}
\ee
where $p=\cos{\chi}$ and $q = \cos{2 \varphi_1}$.
The linear entropy for this state is minimised when $\chi$ and $\varphi_1$ are both integer multiples of $\pi$, which is satisfied when the in-loop unitary $U$ is equal to the identity. In this case the steady state has linear entropy:
\be
   S_L= \frac{1}{2}-\frac{8 c^4 \left(c^2-1\right)^2}{\left(\left(1-2 c^2\right)^2 \lambda-1\right)^2} \, .
\ee
If this CF protocol (`do nothing' in the loop) is applied to a setup with a system and environment of dimension $d$, the steady state will have its largest eigenvalue  (with eigenvector $\ket{0}$) equal to:
\be
    \beta_0 = \frac{4 \left(c^2-1\right) c^2 (d+\lambda -1)+\lambda -1}{d \left(\left(1-2 c^2\right)^2 \lambda -1\right)} \, .
\ee
The other $(d-1)$ eigenvalues are:
\be
   \beta_j= \frac{\left(1-2 c^2\right)^2 (\lambda -1)}{d \left(\left(1-2 c^2\right)^2 \lambda -1\right)} \, .
\ee
For dimension $d$, this steady state will have linear entropy:
\be
    S_L = (1-\frac{1}{d})-\frac{16 c^4 \left(c^2-1\right)^2 (d-1)}{d \left(\left(1-2 c^2\right)^2 \lambda -1\right)^2} \, .
\ee
which is always less than the linear entropy of the maximally mixed state. Note that when $c=\frac{1}{\sqrt{2}}$, $\beta_0=1$ and $\beta_{j\neq0} =0$ and the steady state is pure. This is because, when $c=\frac{1}{\sqrt{2}}$, the partial swap is the square root of the full swap, so applying it twice enacts a full swap, replacing the system mode with the pure environmental mode.
\section{ Measurement-based Feedback for cooling with a clean controller}
We will now look at the effect of our averaged MF protocol when the environment is not maximally mixed. We will consider a qubit system with a generic environmental input state $\eta$ which will have an arbitrary temperature and apply the same MF protocol as in the high temperature case.
After one iteration of our MF protocol, the output state is:
\be 
    \rho_{out} = c^4 \rho_N + s^2 c^2 \eta + s^2 \ket{0}\bra{0} - i s c^3 [\ket{0}\bra{0}, \rho_N] \, .
\ee
Previously, we assumed that $\eta$ was maximally mixed, but it need not be. We will restrict to qubits and choose $\ket{0}$ to point along the direction of the dominant eigenvector of $\eta$, so we can write $\eta = \eta_{0}\ket{0}\bra{0} + (1-\eta_0)\ket{1}\bra{1}$. Solving for steady state gives:
\begin{align}
    \rho_{ss}& = \frac{1}{1-c^4 \lambda}(c^4(1-\lambda)\frac{1}{d}\id + (s^2 c^2 \eta_0 +s^2)\ket{0}\bra{0} + s^2 c^2(1-\eta_0)\ket{1}\bra{1} \\
    & -i \lambda s c^3 [\ket{0}\bra{0}, \rho_{ss}])
\end{align}
As before, this state is diagonal in the basis $\{\ket{j}\}$. It has eigenvalues
\be
    \alpha_0 = \frac{1}{1-c^4 \lambda}\left(c^4 (1-\lambda)\frac{1}{2} + s^2 c^2 \eta_0 + s^2 \right)\, ,
\ee
\be
    \alpha_1 = \frac{1}{1-c^4 \lambda}\left(c^4 (1-\lambda)\frac{1}{2} + s^2 c^2 (1-\eta_0) \right) \, .
\ee
The linear entropy of this state is:
\be \label{mfopteta}
    S_{MF} =\frac{1}{2}-\frac{\left(c^2-1\right)^2 \left(c^2 (2 \eta_0-1)+1\right)^2}{2 \left(c^4
   \lambda-1\right)^2} \, .
\ee
A maximally mixed environment corresponds to $\eta_0 = 1/2$, which recovers the expression from earlier. For a pure environment, we set $\eta_0 = 1$ and we obtain the following expression from the paper:
\be
    S_{MF} = \frac{1}{2} - \frac{(c^4 -1)^2}{2(c^4 \lambda -1)^2} \, .
\ee

We will now prove that, even with a pure environment, no MF protocol can prepare a pure steady state when $c=\frac{1}{\sqrt{2}}$. Recall that, for $c=\frac{1}{\sqrt{2}}$ and a pure environment CF produces a pure steady state, so this is an interesting point of comparison. Again, we will restrict our investigation to qubits. As we have seen before, after projective measurement and the action of a unitary, the system and controller are in a joint state:
\be
    \rho_T=\sum_j p_j \rho_j \otimes \ket{\psi_j}\bra{\psi_j} \, ,
\ee
where 
\be
    \rho_j = \frac{1}{p_j}\big( \frac{1}{2}c^2 \rho_N + s^2 \eta (\rho_N)_{jj} -isc [\ket{j}\bra{j}, \rho_N] \big) \, .
\ee

First, we will show that $\rho_j$ cannot be pure. We do this by writing $\rho_j$ in the basis containing $\ket{j}$ and restrict to the case of interest, when $c=s=\frac{1}{\sqrt{2}}$. This gives us:
\be
    \rho_S^{(1)} = \frac{1}{2}
    \bmat
        1+\rho_{00} & \rho_{01}(1+i) \\
        \rho_{10}(1-i) & 1-\rho_{00}
    \emat
\ee
where $\rho_{ij}$ are the matrix elements of $\rho_N$. This matrix has eigenvalues:
\be
    \lambda_{+/-} = \frac{1}{2}\big(1 \pm \sqrt{\rho_{00}^2 + 2 \rho_{01}\rho_{10}} \big)\geq 0 \, .
\ee
If this state is pure, then one of its eigenvalues will be equal to $0$, and the other equal to $1$. This requires $\sqrt{\rho_{00}^2 + 2 \rho_{01}\rho_{10}} =1$. Since both eigenvalues must be greater than or equal zero, $\rho_{00}^2 + 2 \rho_{01}\rho_{10}\leq1$. The equality is reached only in the case where $\rho_{00}=1$, which could only be the case if $\rho_N$ was pure. However, $\rho_N$ cannot be pure since it has been subject to the depolarising map. Even if $\rho_{in}$ was pure, $\rho_N$ would not be pure for any non-neglible value of the noise parameter $\lambda$. Thus, we can conclude that the $\rho_j$'s are not pure.

After the second system-controller interaction, the controller is traced out and the output state is given by:
\be 
    \rho_{out} = \sum_i p_i \bigg(c^2 \rho_i + s^2 \ket{\psi_i}\bra{\psi_i} -i s c [\ket{\psi_i}\bra{\psi_i}, \rho_i]\bigg) = \sum_i p_i \sigma_i\, .
\ee
By applying the same argument that we used to prove that $\rho_j$ could not be pure, we can prove that $\sigma_j$ also cannot be pure. Thus, since $\rho_{out}$ is a mixture of mixed states, it cannot be pure. This means that no averaged MF protocol can achieve a pure steady state for a partial swap interaction when the coupling is characterised by $c=\frac{1}{\sqrt{2}}$.

\section{Cooling Comparison with intermediate noise}
\begin{figure*}
    \centering
    \subfigure[]{\includegraphics[width=0.45\textwidth]{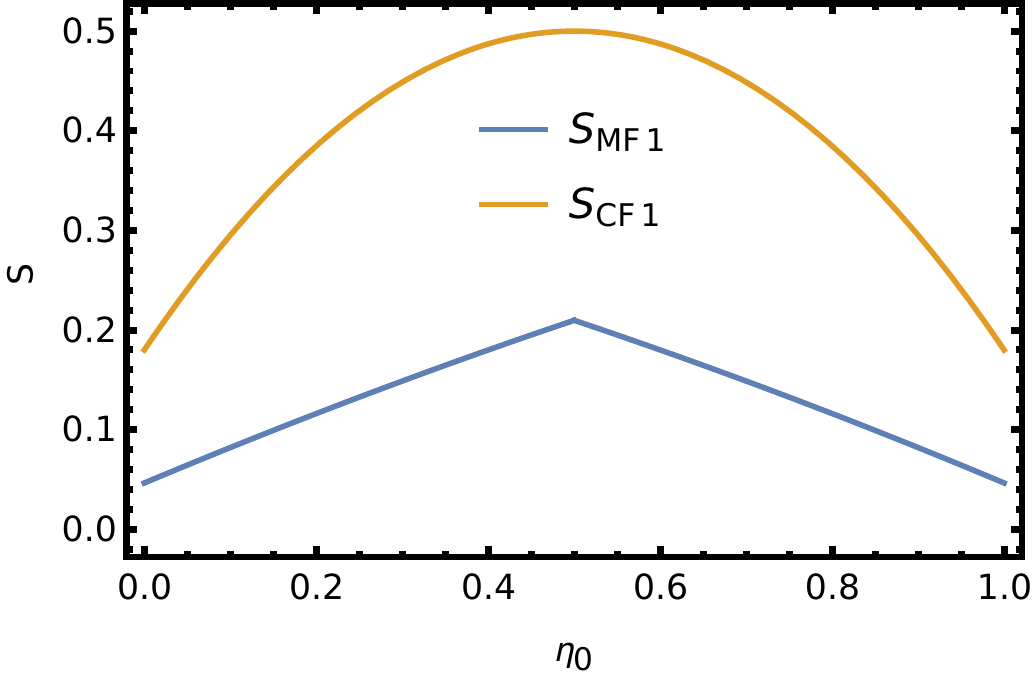}} 
    \hfill
    \subfigure[]{\includegraphics[width=0.45\textwidth]{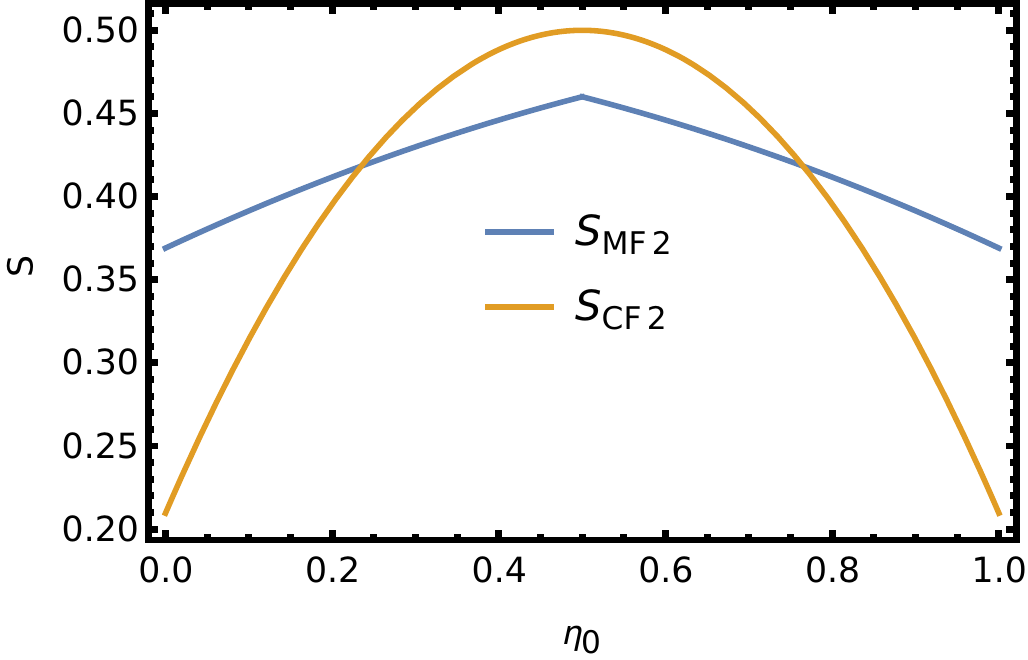}} 
    \caption{(a) Shows the steady state entropy against the largest eigenvalue of the environmental state for MF and CF setups, as given by equations (\ref{mfopteta}) and (\ref{cfopteta}) where $c=0.5$ and $\lambda=0.25$. (b) Shows the same expressions for setups where $c=0.9$ and $\lambda=0.5$.  }
    \label{comparison}
\end{figure*}

We will now apply our CF protocol from earlier, where the in-loop unitary is the identity, to a setup with a non-zero temperature environment with state $\eta = \eta_{0}\ket{0}\bra{0}+(1-\eta_0)\ket{1}\bra{1}$. This leads to a steady state where which is diagonal in the $\{\ket{0}, \ket{1}\}$ basis with the following eigenvalue corresponding to the $\ket{0}$ eigenvector:
\be
    \beta_0=\frac{4 \left(c^2-1\right) c^2 (2 \eta_0 +\lambda -1)+\lambda -1}{2 \left(1-2 c^2\right)^2 \lambda -2} \, .
\ee
This steady state has a linear entropy
\be \label{cfopteta}
    S_{CF} = \frac{1}{2} - \frac{8 c^4 \left(c^2-1\right)^2 (1-2 \eta_0 )^2}{\left(\left(1-2 c^2\right)^2 \lambda -1\right)^2} \, .
\ee
We will now compare the performance MF and CF for intermediate environmental temperatures by comparing equations (\ref{mfopteta}) and (\ref{cfopteta}). Equation (\ref{mfopteta}) gives the steady state system entropy when the MF protocol involves measuring the ancilla and, regardless of the outcome, preparing it in the same state before the system and environment interact for the second time. This equation assumed that, after measurement, the controller is prepared in the state $\ket{0}\bra{0}$. Note that the protocol allows for some flexibility as the controller could be prepared in any state after measurement. It is optimal to use MF to prepare the controller in the state corresponding to the largest eigenvalue of the environmental input. For $\eta_0<1/2$, this corresponds to $\ket{0}\bra{0}$ and for $\eta>1/2$, this corresponds to $\ket{1}\bra{1}$. For our comparison, we assume that the optimal MF protocol is used. The steady state entropy for this protocols, along with steady state CF entropy given by equation (\ref{cfopteta}) is plotted in Fig. \ref{comparison} for different environmental states, as parameterised by $\eta_0$. We find that for the setup with $c=0.5$, MF outperforms CF for all environmental temperatures. For a weaker system-environment interaction, characterised by $c=0.9$ and noise parameter $\lambda=0.5$, we find that MF outperforms CF for high temperature environments with $0.357<\eta_0<0.764$, but CF outperforms MF at low temperatures characterised by $\eta_0<0.357$ and $\eta_0>0.764$. Broadly, we can make the following observation: for low temperature environments and weak couplings, the act of measurement disturbs the coherent process which allows for low entropy environmental states to be transferred to the system, meaning that MF is inferior to CF. However, with strong couplings and noisy environments, the purification from the act of measurement compensates for this and leads MF to be superior to CF.
\section{Coherent Feedback for Protecting an Excited State from Amplitude Damping}
\begin{figure}
    \subfigure[]{\includegraphics[width=0.45\textwidth]{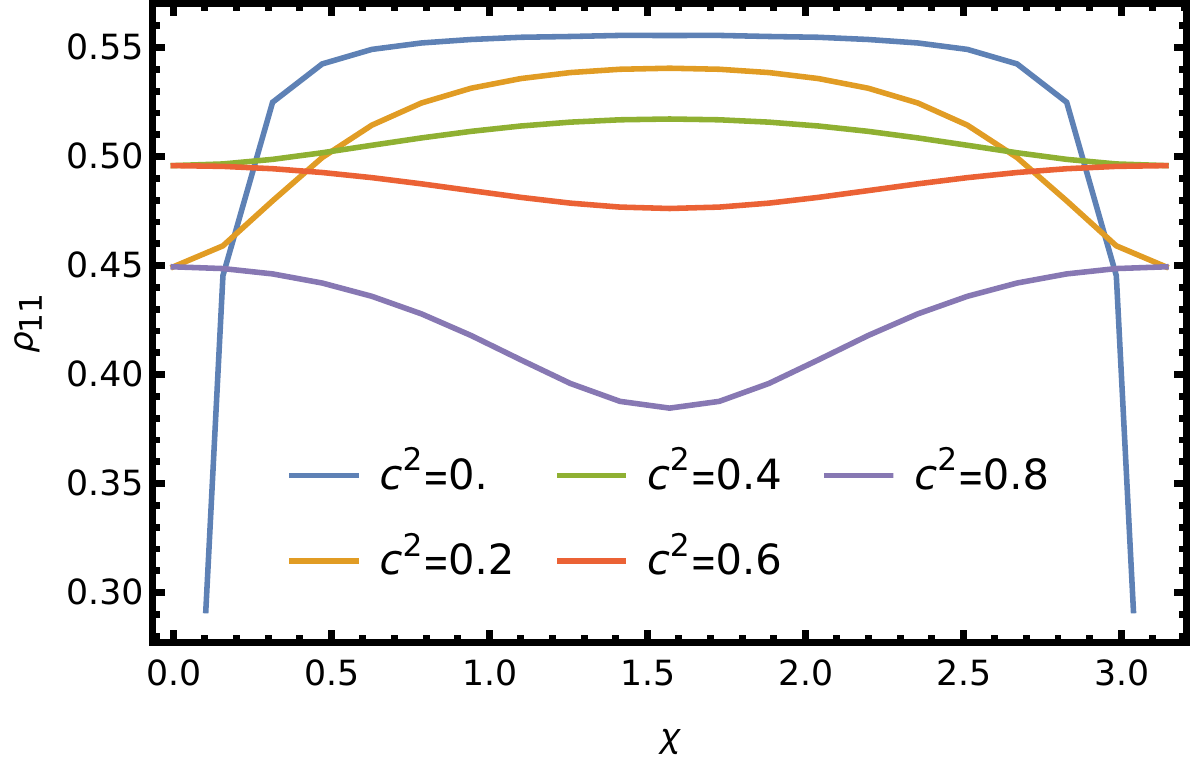} \label{fig:r00vsXf1}}  
    \hfill
    \subfigure[]{\includegraphics[width=0.45\textwidth]{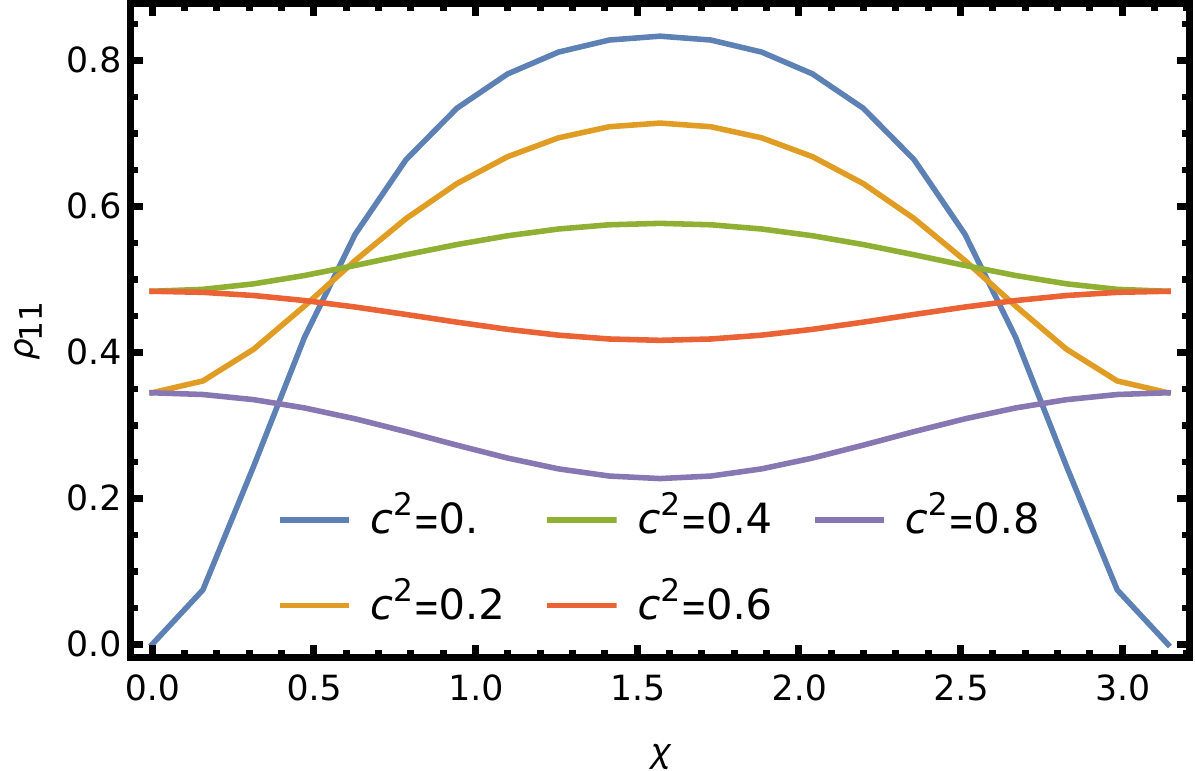} \label{fig:r00vsXf2}} 
    \caption{The steady-state occupation of the $\ket{1}$ state for CF setups with different coupling strengths, characterised by $c^2$. Figure \ref{fig:r00vsXf1} is for a setup with amplitude damping characterised  by $\gamma=0.2$ and Figure \ref{fig:r00vsXf2} is for a setup characterised  by $\gamma=0.8$. Note that, for strong couplings ($c^2<\frac{1}{2}$), the setup with stronger amplitude damping actually has a \emph{higher} excited state occupation.}
    \label{fig:r00vsX}
\end{figure}

We will now investigate the ability of CF to protect a qubit from amplitude damping. The qubit will be subject to amplitude damping channel, after which the CF loop will be applied. We will assume that the highest energy state is the $\ket{1}$ state, so that amplitude damping channel is given by Kraus operators:
\be
    E_0= \sqrt{\gamma}\ket{0}\bra{1} \, , \quad E_1 =\sqrt{1-\gamma}\ket{1}\bra{1} + \ket{0}\bra{0} \, .
\ee
This channel will be applied to the system before the CF loop is applied to attempt to counter it. Using our toy model again, the two system-controller interactions will be partial swaps. We will restrict the in-loop unitaries to rotations of the form:
\be
    U =
    \bmat
        \cos \chi & \sin \chi \\
        -\sin \chi & \cos \chi \\ 
    \emat \, .
\ee
We will assume that the controller is `noisy' and initialised in the maximally mixed state. Our figure of merit will be the steady-state occupation of the $\ket{1}$ state. Investigating numerically, we find that the optimal CF protocol depends on the partial swap strength, characterised by $c$. For $c^2<\frac{1}{2}$, the optimal protocol involves setting $\chi=\frac{\pi}{2}$ and for $c^2>\frac{1}{2}$, the optimal protocol involves setting $\chi=0$. Plots of the steady-state occupation of the $\ket{1}$ state, against $\chi$, for different values of the $c^2$ are plotted in Figure \ref{fig:r00vsX}. 

Plot \ref{fig:r00vsXf1} shows the performance of the setup with weak damping, characterised by $\gamma=0.2$ and plot \ref{fig:r00vsXf2} shows the performance of the same setup with stronger damping, characterised by $\gamma=0.8$. Notice that, for weak system-controller interactions (where $c^2>\frac{1}{2}$), increasing the damping noise strength decreases the steady-state occupation of the excited state, as expected. However, for strong system-controller interactions,(where $c^2<\frac{1}{2}$), the optimal performance (when $\chi=\frac{\pi}{2}$) is actually \emph{improved} by stronger damping. This is because the action of the in-loop $\frac{\pi}{2}$ rotation is more effective at populating the $\ket{1}$ state when more of the state is initially prepared in the $\ket{0}$ state. In this sense, CF allows for the purifying effect of the amplitude damping map to be harnessed for the purpose of increasing the excited state population.
When $\chi=0$ and the in-loop unitary is the identity, the steady-state occupation of the $\ket{1}$ state is:
\be \label{SMAD:X0}
    \rho_{11}^{\chi=0} = \frac{2 c^2 \left(1-c^2\right)}{4 \left(c^2-1\right) (\gamma -1) c^2+\gamma }, 
\ee
which is a decreasing function of $\gamma$ for $c^2>\frac{1}{2}$. Conversely, when $\chi=\frac{\pi}{2}$, the steady-state occupation of the $\ket{1}$ state is:
\be \label{SMAD:Xpi2}
    \rho_{11}^{\chi=\frac{\pi}{2}} = \frac{1-c^2}{2 (\gamma -1) c^2-\gamma +2} \, ,
\ee
which is an increasing function of $\gamma$ for $c^2<\frac{1}{2}$. 
\section{Measurement-based Feedback for Protecting an Excited State from Amplitude Damping}
We will compare this to an intuitive MF protocol which measures the controller in the $\{\ket{1}, \ket{0}\}$ basis, does nothing if the result is $\ket{1}$, and applies an in-loop rotation with $\chi=\frac{\pi}{2}$ if the result is $\ket{0}$, resulting in a controller which is always prepared in the $\ket{1}$ state post-measurement. This protocol results in a steady-state occupation of the $\ket{1}$ state:
\be \label{SMAD:mf}
    \rho_{11}^{MF}=
    \frac{2-c^4-c^2}{2 (\gamma -1) c^4+2} \, .
\ee
This expression is greater than equation (\ref{SMAD:Xpi2}) for all values of $c$ and $\gamma$, so we can say that MF outperforms CF in the regime of $c^2<\frac{1}{2}$. However, for some setups with $c>\frac{1}{2}$, equation (\ref{SMAD:mf}) is slightly lower than (\ref{SMAD:X0}). In particular, (\ref{SMAD:X0}) is greater than (\ref{SMAD:mf}) when:
\be
    c>\frac{1}{2} \sqrt{\frac{-7 \gamma +\sqrt{\gamma  (17 \gamma -24)+16}+4}{2-2 \gamma }} \, .
\ee
In this regime, CF will outperform MF, though numerical investigations suggest that the advantage is small. Conversely, in the regime where MF outperforms CF, the advantage tends to be larger. Figure (\ref{fig:r00vsN}) shows $\rho_{11}$ the occupation of the $\ket{1}$ state, against $N$, the number of iterations of the feedback loop. Both MF and CF protocols described above are presented. In one setup, MF outperforms yields a higher steady-state occupation of the the excited state, and in the other, CF achieves a higher steady-state occupation. The unconditional (averaged) MF trajectory is shown, as well as fifty conditional trajectories for each setup.

\section{Coherent Feedback for performing a bit-flip}
The input for the system will be a pure state given by
\be
    \ket{\psi} = \cos{\frac{\chi}{2}}\ket{0} + e^{i \phi}\sin{\frac{\chi}{2}}\ket{1} \, .
\ee
The environment is initialised to a state of the form $\eta= \eta_0 \ket{0}\bra{0}+(1-\eta_0)\ket{1}\bra{1}$. The system and environment interact though a partial swap unitary given by $U_s = c\id -is\hat{S}$. After this interaction, a $\sigma_x$ unitary is performed on the controller, before the system and controller interact again through $U_s$ and the controller is traced out. For such a setup, the final output state is $\rho_S^{out}$, which does not depend on the value of $\eta_0$ and takes the form:
\be
    \rho_{out}= \frac{1}{2}
\left(
\begin{array}{cc}
 (c^4-s^4) \cos (\chi )+1 & e^{-i \phi } \sin (\chi ) \left(c^2+s^2 e^{2 i
   \phi }\right) \\
 e^{-i \phi } \sin (\chi ) \left(s^2+c^2 e^{2 i \phi }\right) &
   \left(s^4-c^4\right) \cos (\chi )+1 \\
\end{array}
\right)
\ee
The desired final output state is $\rho_X = \ket{\psi_X}\bra{\psi_X}$ where
\be
    \ket{\psi_X} = \sigma_x \ket{\psi} = \cos{\frac{\chi}{2}}\ket{1} + e^{i \phi}\sin{\frac{\chi}{2}}\ket{0} \, .
\ee
The fidelity of the output state to the desired state is a function of $\chi$ and $\phi$ and is given by 
\be
    F_{CF}(\chi, \phi) = \bra{\psi_X}\rho_S^{out}\ket{\psi_X} =\frac{1}{4} (c^2 \left(\cos (2 \phi )-2 \cos (2 \chi ) \cos ^2(\phi )\right)-3 c^2+4) \,.
\ee
As our figure of merit, we take this ouput fidelity, averaged over the Haar measure for the input states, given by:
\be
    A_{CF}=\frac{1}{4 \pi}\int F_{CF}(\chi, \phi) \sin{(\chi)} d \chi d \phi = 1-\frac{2}{3}c^2 \, .
\ee
where $\chi$ is integrated from $0$ to $\pi$ and $\phi$ is integrated from $0$ to $2\pi$.
\section{Measurement-based feedback for performing a bit-flip}
\subsection{Projective Measurements}
Our general projective MF protocol is as follows. The system is initialised in the pure state $\ket{\psi}$ given above, and the controller initialised in the state $\eta= \eta_0 \ket{0}\bra{0}+(1-\eta_0)\ket{1}\bra{1}$. They interact through $U_s$. Then, a measurement is made on the controller. If the result is $\ket{0}$, the unitary $U$ is applied to the controller and if the result is $\ket{1}$, a unitary $V$ is applied instead. This process is equivalent to the POVM with elements $\{U \ket{0}\bra{0}, V\ket{1}\bra{1}\}$ being applied to the controller. After this, the system and controller interact again through $U_s$. The output system state $\rho_{S, MF}^{out}$ is too lengthy to print here, as is the fidelity $F_{MF}=\bra{\psi_X}\rho_{S, MF}^{out}\ket{\psi_X}$. However, after the Haar measure average is taken, we obtain the more reasonable expression:
\be
\begin{aligned}
    A_{MF}&=\frac{1}{4 \pi}  \int F_{MF}(\chi, \phi) \sin{(\chi)} d \chi d \phi  \\ & =\frac{1}{12} (6 - 2 c^4 - s^4 \cos{2 \theta_u} - s^4 \cos{2 \theta_v})  \, ,
\end{aligned}
\ee
where we have used the  decomposition of $2\times 2$ unitary matrices to write magnitudes of the matrix elements $u_{jk}$ and $v_{jk}$ as
\be
    |u_{00}| = \cos{\theta_u} \quad |u_{10}| = \sin{\theta_u}\quad
    |v_{11}| = \cos{\theta_v} \quad |v_{01}| = \sin{\theta_v} \, .
\ee
We find that the Haar measure averaged fidelity is maximised when $\theta_u=\theta_v = \frac{\pi}{2}$, meaning that both $U$ and $V$ are equal to $\sigma_x$ (up to a phase which doesn't affect the outcome). Plugging these optimal unitaries gives the maximum average output fidelity:
\be
    A_{MF}=\frac{1}{3}(1+s^2) = \frac{2}{3}-\frac{1}{3}c^2
\ee
It is straightforward to see that this expression is lower than $A_{CF}$ for all values of $\theta$, except when $c=1$ and there is no feedback present.
\subsection{General POVMs}
In the previous section, we considered MF using projective measurements in-loop, but we can also consider more general POVMs. Here, we will consider the action of general POVMs in between the two system-environment partial swap interactions. The polar decomposition can be used to write the Kraus operators of any POVM, represented with $K_j$ as $K_j = U_j P_j$, where $U_j$ is a unitary matrix and $P_j$ is a positive semidefinite matrix . Note that, to define a POVM, we must have $\sum_j K_j^\dag K_j = \id$ which implies $\sum_k P_j^\dag P_j = \id$. Thus, we can view any general POVM as the action of measurement characterised by $\{P_j\}$, followed by the action of a unitary $U_j$ which depends on the measurement outcome (this observation was made in \cite{jacobs2014coherent}). Since, in MF, we are already allowing for the action of a unitary depending on the measurement outcome, we can absorb $U_j$ into these feedback unitaries and consider the measurement process as entirely characterised by $\{P_j\}$.

Furthermore, since our figure of merit is averaged over the Haar measure, which is unitarily invariant, we can assume that $P_j$ are diagonal in the $\{\ket{0}, \ket{1}\}$ basis. Therefore, the most general qubit POVM can be described using the Kraus operators:
\be
    P_0 = a\ket{0}\bra{0} + b\ket{1}\bra{1} \quad \, , \quad P_1 = \sqrt{1-a^2}\ket{0}\bra{0} + \sqrt{1-b^2}\ket{1}\bra{1} \quad \, .
\ee
For $0\leq a\leq1$ and $0\leq b \leq 1$. Note that when $a=b$, both elements of the POVM are proportional to the identity and correspond to no measurements being performed. When $a=1$ and $b=0$ the POVM corresponds to a projective measurement. Thus, the difference between $a$ and $b$ acts as a measure of the strength of the measurement.

We will absorb the action of $U_j$ into $U$ and $V$, so that the entire process is captured by a POVM with elements $\{U P_0, V P_1\}$. Again, by using the decomposition of $2\times2$ unitary matrices, we come to the optimal protocol, which is when $U=V=\sigma_x$. Applying this optimal protocol yields the following expression for the Haar-measure averaged fidelity:
\be \label{fidPOVM}
    A_{MF}=
    \frac{1}{3} \big( \sqrt{1-a^2}\sqrt{1-b^2} s^2 + a b s^2 +2 -c^2 \big) \, .
\ee
This expression yields the expression previously obtained for projective measurement feeback when $a=1$ and $b=0$. When $a=b=1/2$, we have $P_0=P_1=\frac{1}{2}\id$, which corresponds to no measurement being performed (or `infinitely weak' measurement) and only the action of the unitaries. In this case, (\ref{fidPOVM}) yields the expression previously obtained for coherent feedback. It is in this sense CF can be viewed as MF in the limit of infinitely weak measurements.

It is straightforward to show that the output fidelity (\ref{fidPOVM}) is maximised when $a=b$, thus proving that no measurement-based feedback process can outperform the coherent feedback protocol in this task. However for performing a bit-flip, POVMs corresponding to weaker measurements (where the value of $a$ is closer to $b$) can achieve better performance than stronger measurements, as they disturb the input state less.
\section{Operator Control in the Limit of weak interactions}
We have looked at comparing MF and CF with a partial swap coupling
\be
    U_s^{\theta} = \cos{\theta}\id -i \sin{\theta}\hat{S}
\ee
where $\hat{S}$ is the full swap. When $\theta=0$, both MF and CF are trivially the same, as the system and controller do not interact at all. We will now investigate the effect of MF and CF in the limit of an infinitesimal interaction angle $\theta\xrightarrow[]{}d \theta$.

Expanding $U_s$ to lowest order in $\theta$:
\be
    \lim_{\theta \xrightarrow[]{} d \theta}U_s^\theta = \id -id\theta\hat{S} + o((d \theta)^2) \, .
\ee
Then, to first order we have:
\be
    U_s U_T U_s = U_T -i d\theta \{S, U_T \} \, .
\ee
for any operator $U_T$. The effect of one iteration of CF is
\be
    \rho\xrightarrow{} \tr_C [U_s U_T U_s \rho \otimes \eta U_s^\dag U_T^\dag U_s^\dag] \, ,
\ee
where $U_T=\id\otimes U$ is a unitary which acts only on the controller. Expanding $U_s$ to first order in $\theta$ and discarding higher order terms gives:
\be
    \tr_C\big[U_T \rho\otimes \eta U_T^\dag -i d \theta \{S, U_T\} \rho \otimes \eta U_T^\dag\ +i d \theta U_T \rho\otimes \eta \{S^\dag, U_T^\dag\} \big] \, .
\ee
Using the fact that $\tr_C \big(S A\otimes B=BA \big)$ and the fact that the partial trace is cyclically invariant over the subspace which is being traced over, we obtain:
\be
    \tr_C[\{S, U_T\}\rho\otimes\eta U_T^\dag] = U\eta U^\dag \rho + \eta \rho \, ,
\ee
\be
    \tr_C[U_T\rho\otimes\eta \{S^\dag, U_T^\dag\}] = \rho U\eta U^\dag +\rho \eta \, .
\ee
Combining these, we obtain the transformation:
\be
    \rho_S \xrightarrow[]{} \rho_S+i[\rho_S, U\eta U^\dag] d \theta + i [\rho_S, \eta]d \theta \; .
\ee

To represent more general in-loop operations, we can replace $U$ with a Kraus operator $K_j$ acting on the controller. Due to the linearity of the trace,  we can sum over these Kraus operators and in this way we obtain the expression 
\be
    \rho_S \xrightarrow[]{} \rho_S+i[\rho_S, \Phi(\eta)] d \theta + i [\rho_S, \eta]d \theta \; ,
\ee
where $\Phi(\eta) = \sum_j K_j \eta K_j^\dag$. This expression can be used to describe measurement-based feedback when $K_j$ are used to describe the POVM operators, followed by the action of a unitary.

\end{document}